\documentclass[traditabstract]{aa} 
                                   
\usepackage{amsmath}

\usepackage{graphicx}
\usepackage{grffile}

\usepackage{txfonts}
\usepackage[round]{natbib}
\bibpunct{(}{)}{;}{a}{}{,} 

\usepackage[bookmarks]{hyperref}
\usepackage{color}

\begin{document}
   \title{Thermal and Photophoretic Properties of Dust Mantled Chondrules and Sorting in the Solar Nebula}

   \author{C. Loesche
          \inst{1} \and
          G. Wurm
          \inst{1}
          }

   \institute{Faculty of physics, University of Duisburg-Essen,
              Lotharstr. 1, D-47057 Duisburg\\
              \email{christoph.loesche@uni-due.de}
             }

   \date{June 6, 2012}
 
  \abstract
   {Many chondrules are found to be surrounded by a fine grained rim. Supposedly, these rims were acquired from dust accreted to the chondrules in a protoplanetary disk. In numerical simulations we study the heat transfer in illuminated bare and dust mantled chondrules. The calculations consider the chondrule size, the dust mantle size, and the thermal conductivities of both components as parameters. We calculate the photophoretic force and compare the numerical results to analytical approximations. We give an expression to quantify the photophoretic force on a spherical particle in the free molecular regime to better than 2\%. We describe the influence of a dust mantle on the photophoretic strength by an effective thermal conductivity of the core-mantle particle. The effective thermal conductivity significantly depends on the size ratio between mantle and chondrule but not on the absolute sizes. It also strongly depends on the thermal conductivity of the mantle with minor influence of the thermal conductivity of the chondrule. The size ratio between rim and chondrule in meteorites is found to vary systematically with overall size by other authors. Based on this, our calculations show that a photophoretic size sorting can occur for dust mantled chondrules in optically thin disks or at the evolving inner edge of the solar nebula.}

   \maketitle

\section{Introduction}

The work behind this paper is based on two observations concerning the mm-size particles, called chondrules, which are a major component in primitive meteorites. 
\begin{itemize}
	\item Chondrules often have specific but different size in different meteorites \citep{Kuebler1999, Scott1996,  Hughes1978}.
	\item Chondrules are often surrounded by fine grained rims \citep{Metzler1991}.
\end{itemize}
The origin of these facts is still debated. The rims might originate in the solar nebula. As bare chondrules move through a dusty disk they would collect the dust. This dust is compressed to the observed rims, eventually
\citep{Bland2011, Metzler1992, Ormel2008, Morfill1998, Carballido2011}. From an experimentalists point of view this is very plausible as it is almost impossible to prevent that a chondrule in a dust rich environment will enshroud itself in a mantle of small grains \citep{Blum2008}. Specific experiments to study accretion of dust rims and further evolution have recently been carried out by \citet{Beitz2011}.
As far as the size sorting is concerned also a number of different scenarios are sketched in the literature. The local reservoirs of matter in the solar nebula, which created chondrules and which fed a parent body, might just have produced different size distributions in different locations. More often though it is discussed that certain mechanisms might have acted on a particle reservoir, which separated chondrules of different size \citep{Liffman2005}. Aerodynamic aspects are clearly tempting here as particles of different size couple differently to the embedding gas \citep{Cuzzi1996}. All the explanations for size sorting and rim formation seem to have their advantages and shortcomings. We cannot dismiss one or the other as within their set of assumptions all might currently be viable scenarios. In this paper we assume that the rims set off as highly porous dust mantles on chondrules of nebula origin and add just one more assumption, namely that the particles are illuminated by the young sun. Adding a light source might then naturally lead to a (size) sorting due to photophoresis.

Photophoresis was first introduced to the physics of protoplanetary disks by \citet{Krauss2005}. First attempts to explain size sorting and concentration of chondrules by photophoresis were made by \citet{Wurm2006}. They showed that in an optically thin solar nebula chondrule aggregates are sorted due to the size of the constituents in different locations. Key factor here is the thermal conductivity of the aggregates which determines the temperature gradient upon illumination. First experiments to measure photophoresis on chondrules have been carried out by \citet{Wurm2010}. The measurements support the idea that photophoresis acts sufficiently strong on chondrules to provide sorting. 
The assumption of a directed radiation source is vital to the problem and has two very clear sides. On one side, without a directed radiation source (visible or infrared) photophoresis does not exist. Therefore, photophoresis does not exist in the midst of an optical thick protoplanetary disk. However, on the other side, given a light source it is also inevitable that photophoresis acts as significant force on the particles. In the inner parts of an optical thin protoplanetary disks it is orders of magnitudes larger than a stars gravity and cannot be neglected \citep{Moudens2011, Krauss2007, Mousis2007, Wurm2009, Takeuchi2008, Herrmann2007}. 
The assumption that the solar nebula was optically thin at certain times and places is supported by observations of protoplanetary disks. There are a number of disks observed now which have large inner clearings, which are optically thin \citep{Calvet2002, Sicilia-Aguilar2008}. In some of these disks the star is still accreting matter so that the inner region still must contain gas. Clearly the conditions for photophoresis are given here. Also, disks in general have an inner edge. At the inner edge of a disk sorting of chondrules is possible as some particles might be pushed into the dark a bit stronger than others. This idea has been suggested by \citet{Haack2007} and \citet{Wurm2006a}. To evaluate the potential of size sorting and photophoretic transport in further detail in the future we carried out numerical heat transfer calculations on bare chondrules and dust mantled chondrules as prerequisite here. These are used to quantify the photophoretic strengths on particles more accurately.
The structure of this paper is as follows. Section 2 describes our heat transfer simulations, which result in temperature fields at the surface of a (dust mantled) chondrule that is illuminated. In section 3 the resulting temperature field at the surface of a particle is used to calculate photophoretic forces. These forces are compared to values, which are calculated from common equations found in the literature. To facilitate analytical treatment of photophoretic forces in future work we introduce  correction factors on the order of 20\% for homogeneous particles of given size and thermal conductivity. These results are then used in section 4 to define average thermal conductivities for dust mantled chondrules depending on the chondrule size, dust mantle size and the thermal conductivities of both components. In section 5 we sketch how chondrules might be size sorted due to photophoresis based on our calculations.

\section{Heat transfer calculations}

\begin{figure}[h]
\begin{center}
\includegraphics[width=\columnwidth]{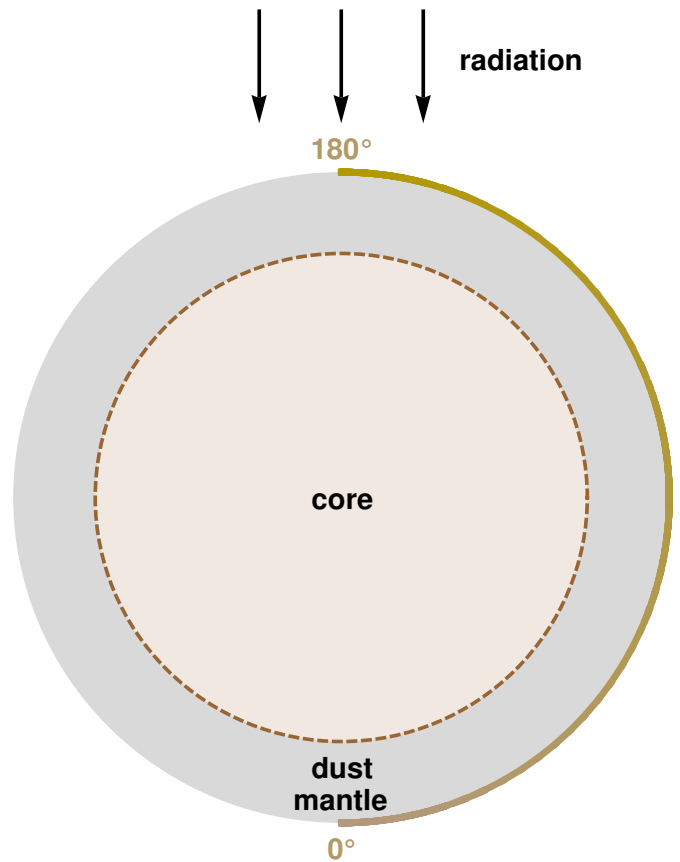}
\caption{Model setup of a chondrule core (high thermal conductivity) and a dust mantle (low thermal conductivity). The light source is directed downwards. Due to rotational symmetry only the temperature distribution along one line from top ($180\,^{\circ}$) to the bottom ($0\,^{\circ}$) has to be considered.}
\label{fig1}
\end{center}
\end{figure}

Our basic model setup consists of a spherical particle, which is made up of a core (bare chondrule) and a dust mantle (rim) (fig. \ref{fig1}). 

\begin{figure}[h]
\begin{center}
\includegraphics[width=\columnwidth]{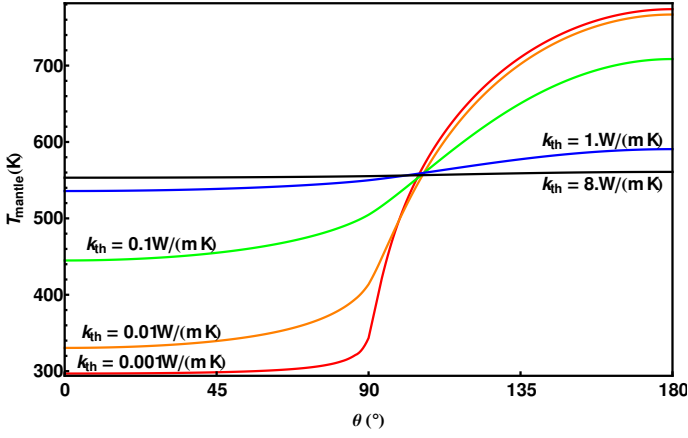}
\caption{Temperature distributions for homogeneous spheres of $1.12\,\mathrm{mm}$ diameter with varying thermal conductivities.}
\label{fig2}
\end{center}
\end{figure}

Both components have specific thermal conductivities. The core, being compact and consisting of silicates, has a thermal conductivity on the order of $k_{core} = 1\,\mathrm{W/(m\,K)}$  \citep{Opeil2010}. The dust mantle being highly porous has a very low thermal conductivity. We vary $k_{dust}$ between $0.01\,\mathrm{W/(m\,K)}$ and $0.5\,\mathrm{W/(m\,K)}$ \citep{Presley1997, Presley1997a, Krause2011}. This core-mantle sphere is illuminated from the top (fig. \ref{fig1}), absorbs radiation and heats up until in equilibrium it emits as much thermal radiation as it absorbs visible radiation. Due to the directed illumination a temperature gradient over the particle surface is established. The problem to solve is the heat transfer equation

\begin{equation}
	\boldsymbol{\nabla} \cdot k\boldsymbol{\nabla} T = Q \; , \label{heatEq}
\end{equation}

where $k$ is the local thermal conductivity, $T$ is the local temperature and $Q$ is a volumetric heat source. As boundary condition we have cooling due to thermal emission and heating due to illumination. We consider a complete absorption of incident light at the surface (i.e. $Q=0$). We further assume an emissivity for thermal radiation of $1$. For a plane wave light source of light flux $I*\mathbf{e}_I$ we then get a von Neumann boundary condition 

\begin{equation}
k 	\boldsymbol{\nabla}T \cdot \mathbf{n} = I\,\mathbf{e}_I \cdot \mathbf{n}-\sigma(T^4-T_{gas}^4) \; , \label{boundaryCond}
\end{equation}

where $\mathbf{n}$ is normal to the surface pointing outwards and the first term on the right-hand side is only considered for the upper illuminated half sphere of the particle.
The light flux in all calculations is fixed at $I = 20\,\mathrm{kW/m^2}$. Eq. \ref{heatEq} and \ref{boundaryCond} are solved numerically using COMSOL, which is based on a finite element method to solve differential equations (COMSOL  v4.1 from COMSOL  AB, Sweden). In case of no rotation the temperature distribution is symmetric around the axis given by $\mathbf{e}_I$. The maximum temperature $T_{max}$ is at the top surface element facing the light source (at $180\,^{\circ}$). The minimum temperature $T_{min}$ is at the opposite side (bottom or $0\,^{\circ}$). The environment is assumed to have $T_{gas} = 293\,\mathrm{K}$. This is far below the equilibrium temperature for an illuminated particle but reflects experimental settings, e.g. in \citet{Wurm2010}. It will also allow us to specify dependencies on temperatures more detailed and not just refer to a single average gas temperature, i.e. we will not assume one average temperature for the gaseous environment and the particle. 
A sample of calculations is shown in Fig. \ref{fig2} for a homogeneous $1.12\,\mathrm{mm}$ particle (total diameter). To calculate homogeneous particles we use the same thermal conductivity for the dust rim and the chondrule (core).

All areas of validity and restrictions of the model are discussed in Sec. \ref{caveats}.

\section{Photophoretic forces}
\subsection{general case at low gas pressure}
The photophoretic force depends on the temperature distribution over the surface of the chondrule. We consider the case of free molecular flow or low gas pressure here, where the mean free path of the gas molecules is much larger than the particle size. In that case the interaction between particle and gas can be treated as individual collisions of the gas molecules with the surface. A gas molecule impinging the surface of the particle is either reflected specular or diffusely. In the first case, due to the larger mass of the solid particle the absolute momentum of the gas molecule does not change and does not depend on the temperature of the particle. If this first fraction of particles, $1-\alpha$, is the same over the surface of the particle, then due to symmetry, specular reflected gas molecules on average do not transfer momentum to the particle. The second fraction of gas molecules, $\alpha$, which are reflected diffusely, stick to the surface for a short while, accommodate to the surface temperature and the momentum of the reemitted gas molecules differs from the momentum of the incoming gas molecules. On a hot surface part the gas molecules leave faster than on colder parts. Therefore the gas carries a net momentum after the interactions. The total force on the particle is the integral over the total particle surface of the local pressure induced by the interaction at the local temperature and in the free molecular flow regime is given as \citep{Rohatschek1985, Rohatschek1995, Hidy1970}

\begin{equation}
	\mathbf{F}=-\frac{1}{2} \oint\limits_{S_{particle}} p \left( 1+\sqrt{\frac{T'}{T_{gas}}} \right) \mathbf{d}\boldsymbol{\sigma} \; , \label{force}
\end{equation}

where

\begin{displaymath}
	T'=T_{gas}+\alpha\left(T-T_{gas}  \right)
\end{displaymath}

denotes the statistical mean temperature and $\alpha$ is the thermal accommodation coefficient describing the fraction of gas particles being adsorbed by the solid particle's surface, $T$ is the local temperature of the particle surface, $T_{gas}$ and $p$ are the gas temperature and pressure far away from the particle, respectively. As $T_{gas}$ in our calculations we use $293\,\mathrm{K}$. 

\subsection{rotational-symmetric temperature field}
In case of a spherical rotational-symmetric temperature field $T=T(\theta)$, e.g. if a sphere is illuminated with an intensity $I$ in direction $\mathbf{e}_I$, without loss of generality let $\mathbf{e}_I:= -\mathbf{e}_z$ (see fig. \ref{fig1}) in order to develop $T$ in Legendre Polynomials  $P_n$:
\begin{equation}
T(\theta)=\sum_{n}A_n \, P_n\left(\cos (\theta) \right)\;,
\end{equation}
with $A_i$ as the $i^{th}$ expansion coefficient.
\citet{Rohatschek1995} further linearizes the square root to its first order.
Due to the Legendre Polynomial's orthogonality relation $\mathbf{F}$ (see eq.  \ref{force}) is only a function of $A_1$; all other $A_i$  do not show up since the corresponding integrals yield zero. Therfore, eq. \ref{force} yields \citep{Rohatschek1995}

\begin{equation}
	\mathbf{F}\simeq-\frac{\pi}{3} \alpha  \frac{p}{T_{gas}}  r^2 A_1  \mathbf{e}_z \; , \label{force1}
\end{equation}
where $r$ denotes the sphere's radius.

Since $A_1$ is usually unknown there are three well known approximations for this coefficient. The first one is (indication ${(i)}$ refers to the $i^{th}$ approximation presented)
\begin{equation}
	A_1^{(1)}\simeq\frac{1}{2} \Delta T \;,\label{koeff1}
\end{equation}

with $\Delta T$ being the maximum temperature difference between top and bottom $\Delta T = T_{\max} - T_{\min}$. Thus eq. \ref{force1} yields an approximated force $\tilde{F}^{(1)}$ for a non-rotating illuminated sphere:

\begin{equation}
	\tilde{F}^{(1)}=\frac{\pi}{6} \alpha  \frac{p}{T_{gas}}   r^2 \Delta T \; .	\label{forceA1}					
\end{equation}

Our first goal here is to calculate the photophoretic force for a homogeneous sphere (core and rim share the same thermal conductivity $k$) based on eq. \ref{force} exactly but describe the results in a way similar to the simple analytic expression of eq. \ref{forceA1}. In the following we therefore determine a correction factor from simulations. We define as correction
	
\begin{equation}
	z^{(1)}:=\frac{F}{\tilde{F}^{(1)}} \; . 
\end{equation}

\begin{figure}[h]
\begin{center}
\includegraphics[width=\columnwidth]{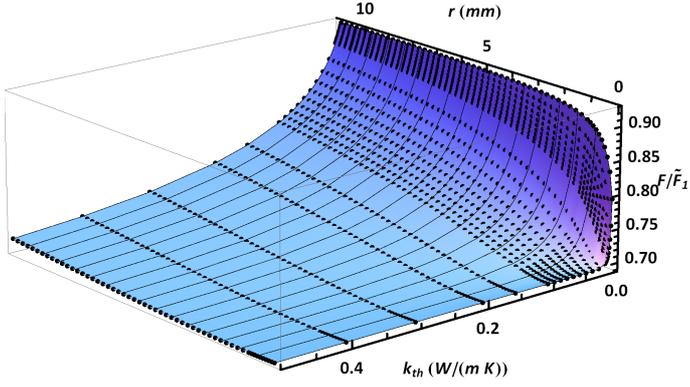}
\caption{Ratio $z^{(1)}=\frac{F}{\tilde{F}^{(1)}}$ for a given particle size and thermal conductivity $k$. For thermal conductivities larger than $k = 0.2$ the factor is essentially constant $z^{(1)} = 0.7$ ($\alpha = 1$, this plot) (resp. $z^{(1)} =0.8$ for $\alpha = 0.8$).}
\label{fig3}
\end{center}
\end{figure}

Varying size $r$ and thermal conductivity $k$ we find that in general, the approximation overestimates the photophoretic force. The correction factor can be considered a constant of $z^{(1)} = 0.7$ for thermal conductivities larger than about $k = 0.2\,\mathrm{W/m\,K}$ independent of the particle size (fig. \ref{fig3}). For smaller thermal conductivities there is a dependency of $z^{(1)}$ on both parameters. As the thermal conductivity of dust falls in the low range a non-constant correction factor has to be considered. The parameters studied range from $r = 0.1 \ldots 11\,\mathrm{mm}$, $k = 0.01 \ldots 8\, \mathrm{W/m\,K}$, and $\alpha = 0.1 \ldots 1$, the latter is explicitly part of eq. \ref{force1}. To give an analytic equation we tried different simple expressions. One expression that fits the correction factor over the whole range with an accuracy being better than 2\% is

\begin{equation}
	z^{(1)}=0.740-0.209 e^{-0.00435 \frac{r}{k}}+0.478 e^{-1.04 \alpha} \; . \label{korr1}
\end{equation}
It should be noted, that in eq. \ref{korr1}, \ref{korr2}, \ref{korr3} and \ref{forceNumerical} $r$ is in units of $\mathrm{mm}$ and $k$ in units of W/(m K).

The next step is expressing the temperature difference in terms of an illuminating flux. For a homogeneous sphere of thermal conductivity $k$ \citet{Rohatschek1995} finds

\begin{equation}
	A_1^{(2)} \simeq \frac{1}{2}\frac{I\, r}{k} \; , \label{koeff2}	 
\end{equation}

where $I$ is the radiative flux. As \citet{Rohatschek1995} notes, this is an approximation for ‘good’ conductors. With low thermal conductivity dust mantles on high conductivity chondrules, it is not clear that the approximation of eq. \ref{koeff2} is sufficient to study the details of photophoretic particle transport and sorting. However, eq. \ref{force1} and eq. \ref{koeff2} are useful as they provide a rather simple analytical treatment of photophoresis including all important quantities, i.e. the gas properties $p$, $T_{gas}$, the radiative flux $I$ and accommodation coefficient $\alpha$, and the particle properties $r$, $k$. To keep this simplicity but to be accurate enough to apply the equations to problems of photophoretic transport of chondrules we chose again a correction term. We extend our correction factor including deviations introduced by the approximation of eq. \ref{koeff2} and define 
	
\begin{equation}
	z^{(2)}=\frac{F}{\tilde{F}^{(2)}}
\end{equation}

with

\begin{equation}
	\tilde{F}^{(2)}=\frac{\pi}{6} \alpha  \frac{p}{T_{gas}}  r^3 \frac{I}{k} \; .\label{forceA2}
\end{equation}

\begin{figure}[h]
\begin{center}
\includegraphics[width=\columnwidth]{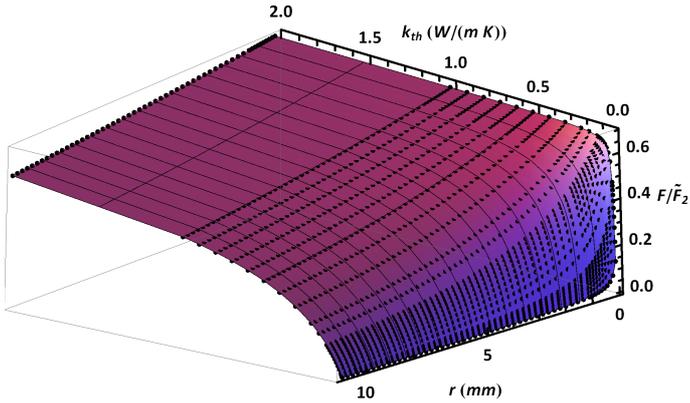}
\caption{Ratio $z^{(2)}=\frac{F}{\tilde{F}^{(2)}}$ for a given particle size and thermal conductivity $k$. For thermal conductivities larger than 2 the factor is essentially constant. The accommodation coefficient $\alpha$ is $1$.}
\label{fig4}
\end{center}
\end{figure}

As before, the approximation (eq. \ref{forceA2}) overestimates the photophoretic force following the same pattern as before. In the same parameter set ($r = 0.1 \ldots 11\,\mathrm{mm}$, $k = 0.01 \ldots 8\,\mathrm{W/m\,K}$, $\alpha = 0.1\ldots 1$) $z^{(2)}$ can be well fitted again to better than 2\% by the following expression

\begin{equation}
	z^{(2)}=0.590-0.0599 e^{-0.0359 \frac{r}{k}}+0.449 e^{-0.997\alpha} \; . \label{korr2}
\end{equation}
 
This correction factor (fig. \ref{fig4}) can be rather far from unity especially for low thermal conductivity particles. Clearly, particles with thermal conductivities much lower than $1\,\mathrm{W/m\,K}$ can no longer be regarded as good conductors in the sense of eq. \ref{koeff2}. Doing so would usually overestimate the photophoretic force significantly. This is due to the fact that the approximation of eq. \ref{koeff2} neglects the effect of thermal radiation. This aspect can be improved by adding a term to eq. \ref{koeff2} which accounts for this (e.g. cmp. \citet{Beresnev1993})

\begin{equation}
	A_1^{(3)}\simeq\frac{1}{2} \frac{I}{\frac{k}{r}+4 \sigma T_{par}^3} \; . \label{koeff3}
\end{equation}

Usually, particles are considered in thermal equilibrium with their surroundings, photophoretic temperature gradients are small and illumination is not significantly changing the average temperature. In that situation the particle temperature $T_{par}$ is approximated by $T_{gas}$. For our calculations and often for experimental settings this is not the case, i.e. the particle temperature can easily be a factor $2$ higher than the surrounding gas temperature. As the radiative cooling is depending on the particle temperature it is the particle temperature and not the gas temperature, which enters in eq. \ref{koeff3}. As seen in fig. \ref{fig2} the particle temperature varies strongly over its surface and is still not well defined. We use the average black body temperature here and define

\begin{equation}
	T_{par}=\sqrt[4]{\frac{I}{4\sigma}+T_{gas}^4}	\label{blackBodyTemp} \; .	
\end{equation}

Based on eq. \ref{koeff3} and \ref{blackBodyTemp} we finally consider as analytical approximation that captures all physical parameters of the system but is accurate enough to treat dust mantled chondrules

\begin{equation}
	\tilde{F}^{(3)}=\frac{\pi}{6} \alpha  \frac{p}{T_{gas}}  \frac{r^2  I}{\frac{k}{r}+4\sigma\left(\frac{I}{4\sigma}+T_{gas}^4 \right)^{\frac{3}{4}}}  	\label{forceA3}	
\end{equation}

\begin{figure}[h]
\begin{center}
\includegraphics[width=\columnwidth]{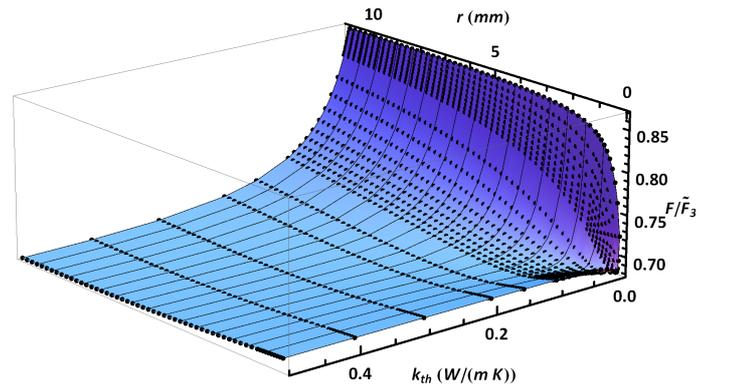}
\caption{Ratio $z^{(3)}=\frac{F}{\tilde{F}^{(3)}}$ for a given particle radius and thermal conductivity $k$. For thermal conductivities larger than $0.2$ the factor can be regarded as constant. The accommodation coefficient $\alpha$ is $1$.}
\label{fig5}
\end{center}
\end{figure}

and quantify the photophoretic force adding a correction factor $z^{(3)}$

\begin{equation}
	F=z^{(3)}\, \tilde{F}^{(3)} \; .
\end{equation}

Similar in structure, $z^{(3)}$, can well be approximated in the whole range of parameters to better than about 2\% by the following terms

\begin{equation}
	z^{(3)}=0.725-0.177 e^{-0.00205  \frac{r}{k}}+0.434 e^{-1.14 \alpha} \; . \label{korr3}
\end{equation}

As summary, if the temperature gradient is known the first approximation (eq. \ref{forceA1} and \ref{korr1}) might be used. More general and used further in this paper is the photophoretic force on a homogeneous, perfectly absorbing spherical particle as given by eq. \ref{forceA3} to \ref{korr3} or combined as

\begin{equation}
\begin{split}
F=\left(0.380-0.0927 e^{-0.00205  \frac{r}{k}}+0.227 e^{-1.14 \alpha} \right)*\\ *\, \alpha \frac{p}{T_{gas}}  r^2  \frac{I}{\frac{k}{r}+4\sigma\left(\frac{I}{4\sigma}+T_{gas}^ 4\right)^{\frac{3}{4}}} \; . \label{forceNumerical}
\end{split}
\end{equation}

\section{Dust mantled chondrules}

In view of equation \ref{forceNumerical}, thermal conductivity of homogeneous spheres is a major particle property that determines photophoresis. For dust mantled chondrules there are two distinct different thermal conductivities for the core and the mantle and it is not clear, a priori, what determines the photophoretic strength of such particles. A core mantle particle can be described by five parameters, which we vary within the following range: 

\begin{itemize}
	\item the core radius,\\ $r_{core}$: $0.2 \ldots 1.0\,\mathrm{mm}$,
	\item the mantle thickness,\\ $d_{dust}$: $0.1 \ldots 1$ in fractions of $r_{core}$ \\(the total radius then is $\rho:=r_{core}\left(1+d_{dust} \right)$),
	\item the thermal conductivity of the core,\\ $k_{core}$: $1 \ldots 4\,\mathrm{W/m\,K}$,
	\item the thermal conductivity of the mantle,\\ $k_{dust}$: $0.01 \ldots 0.5\,\mathrm{W/m\,K}$,
	\item and the accommodation coefficient,\\ $\alpha$: $0.1 \ldots 1$.
\end{itemize}

For each combination of these parameters a photophoretic strength can be calculated. In principle we could argue on the photophoretic strength depending on these parameters at this point. We note that different configurations $\gamma=\left(r_{core},d_{dust},k_{core},k_{dust}\right)$ of dust coated chondrules might result into the same photophoretic strength, even if the temperature field across the mantle is absolutely different. So two configurations $\gamma_1 , \gamma_2$ are related to each other if
\begin{equation}
\gamma_1 \sim \gamma_2 :\Leftrightarrow F(\gamma_1 )=F(\gamma_2) \; .\label{equivalenceRelation}
\end{equation}
Nevertheless, it is possible to choose a representative $\gamma\left(\rho, \kappa \right)$ of all those configurations obeying eq. \ref{equivalenceRelation} with a (total) radius $\rho$ and constant thermal conductivity $\kappa$.
%
We call $\kappa$ effective thermal conductivity.

\begin{figure}[h]
\begin{center}
\includegraphics[width=\columnwidth]{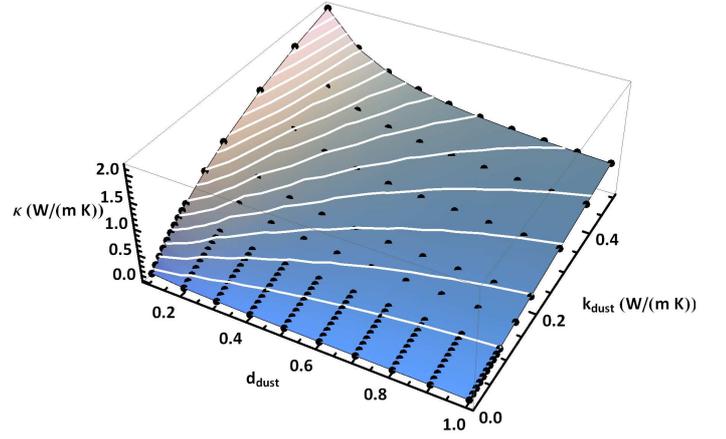}
\caption{$\kappa$ over $d_{dust}$ and $k_{dust}$ for a $1\,\mathrm{mm}$ sized core with a thermal conductivity of $4\,\mathrm{W/m\,K}$ . The accommodation coefficient $\alpha$ is $1$; white lines indicate constant values and show multiple configurations can lead to the same $\kappa$.}
\label{fig8}
\end{center}
\end{figure}

In view of quantifying thermal properties of dust mantled spherical chondrules with a total radius  $\rho\left(r_{core}, d_{dust} \right)  = r_{core} \left(1+d_{dust}\right)$  we use the results of the photophoretic strength calculations and eq. \ref{forceA3} and \ref{korr3} to calculate $\kappa\left(r_{core},d_{dust},k_{core},k_{dust}\right)$. It is of course not injective, so multiple configurations can lead to a same $\kappa$ (see fig. \ref{fig8}). In the following the dependency of  $\kappa$ on the different parameters is considered.
We therefore numerically solved

\begin{equation}
	F\left(T,\alpha\right)=z^{(3)} \left(\rho,\kappa,\alpha\right)\, \tilde{F}^{(3)} \left(\rho,\kappa,\alpha\right)
\end{equation}

for $\kappa$, where $T=T\left(r_{core},d_{dust},k_{core},k_{dust}\right)$ is the sphere's surface temperature (depending on all four parameters for the heat equation system).

\textbf{Accommodation coefficient:} The accommodation coefficient is not connected to $\kappa$ (by construction). Numerically, variations only influence the result on the 1\% level. This is in agreement with the idea that thermal conductivity is independent of the photophoretic property.

\begin{figure}[h]
\begin{center}
\includegraphics[width=\columnwidth]{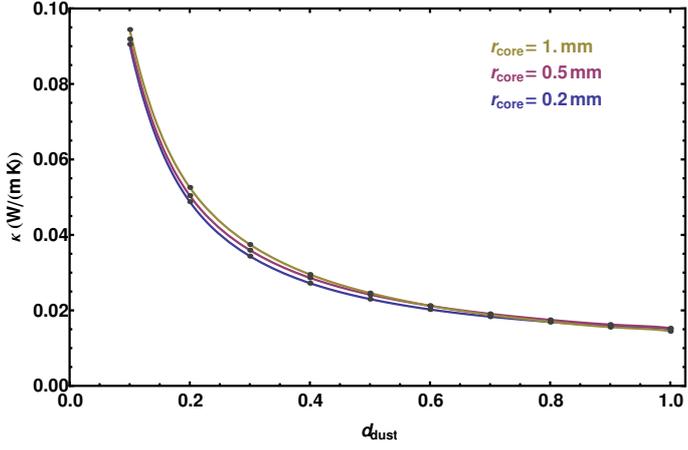}
\caption{Effective thermal conductivity $\kappa$ over dust mantle to core size ratio. The 3 different lines represent 3 different core sizes: from bottom to top $r_{core} = 0.2 , 0.5 , 1.0\,\mathrm{mm}$. The thermal conductivity of the dust mantle is $0.01\,\mathrm{W/m\,K}$, the core's thermal conductivity is $1\,\mathrm{W/m\,K}$  and the accommodation coefficient $\alpha$ is $1$.}
\label{fig6}
\end{center}
\end{figure}

\textbf{Core and mantle size:} In view of estimating the potential for size sorting the most important dependencies to explore are the influence of the total size of the chondrule and the influence on the ratio between mantle thickness and core radius. We find that only for low thermal conductivities of the dust mantle does the effective thermal conductivity slightly depend on the total size. The extreme for a dust mantle of the lowest thermal conductivity studied of $k_{dust}=0.01\,\mathrm{W/m\,K}$ is seen in fig. \ref{fig6}. The thermal conductivity of the core is $1\,\mathrm{W/m\,K}$. All three lines vary very little. This shows that the dependency on total size for a given mantle to core size ratio is very weak. For given thermal conductivities of core and mantle only the ratio between the sizes of both components determines the effective thermal conductivity. This trend can be understood as the smaller the mantle the more effective is the high thermal conductivity of the core. In the extreme of a vanishing mantle the thermal conductivity has to reach the core value, which is consistently confirmed by the calculations. For thick mantles the low thermal conductivity of the mantle should dominate, which is observed in fig. \ref{fig6}. However, it should be noted that even for equal size components the effective thermal conductivity is still a factor 2 larger than the thermal conductivity of the pure dust mantle. 

\begin{figure}[h]
\begin{center}
\includegraphics[width=\columnwidth]{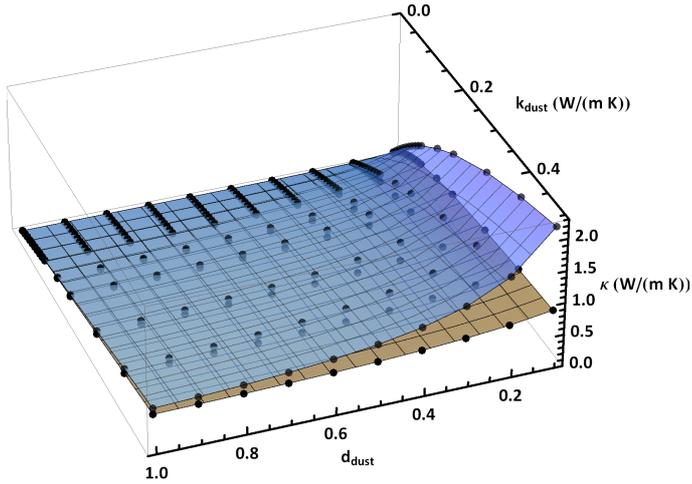}
\caption{$\kappa$ over $d_{dust}$ and $k_{dust}$ for a $1\,\mathrm{mm}$ sized core with a thermal conductivity of $1\,\mathrm{W/m\,K}$ (bottom) and $4\,\mathrm{W/m\,K}$ (top). The accommodation coefficient $\alpha$ is $1$.}
\label{fig7}
\end{center}
\end{figure}

\textbf{Thermal conductivity of core and mantle:} From fig. \ref{fig6} it is obvious that the thermal conductivity of the mantle is a dominating factor for the effective thermal conductivity. In fig. \ref{fig7} the dependency on the size ratio between rim and core and the thermal conductivity of the mantle is shown for two core thermal conductivities. The two different plots are based on the extreme values used for the thermal conductivity of the core in our calculations. The comparison illustrates that the core has a significant influence only for thin mantles and then only if these mantles have high thermal conductivities. 

\begin{figure}[h]
\begin{center}
\includegraphics[width=\columnwidth]{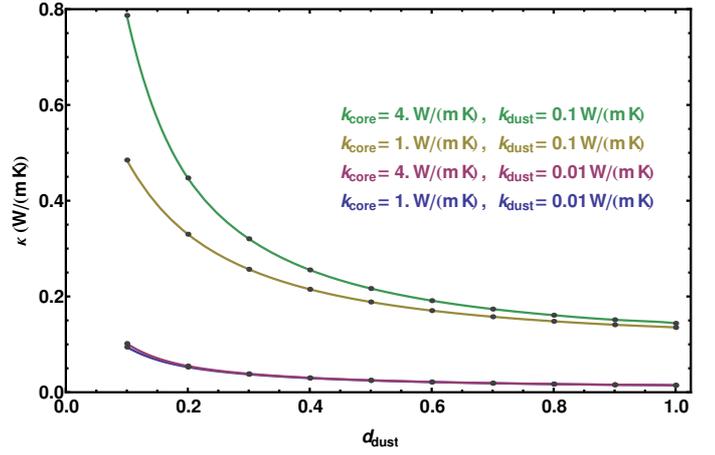}
\caption{$\kappa$ for different pairs $\left(k_{core},k_{dust}\right)$. Two bottom lines: $k_{dust}=0.01\,\mathrm{W/m\,K}$, two top lines: $k_{dust}=0.1\,\mathrm{W/m\,K}$}
\label{fig9}
\end{center}
\end{figure}

Fig. \ref{fig9} shows the dependency in higher resolution again by plotting $\kappa$ over $d_{dust}$ but for two fixed thermal conductivities of both the mantle ($0.01\,\mathrm{W/m\,K}$, $0.1\,\mathrm{W/m\,K}$) and core ($1\,\mathrm{W/m\,K}$, $4\,\mathrm{W/m\,K}$). Even for thin dust mantles, if the thermal conductivity has a reasonable value attributed to dust, then the effective thermal conductivity does not significantly depend on the thermal conductivity of the core.

\textbf{Conclusion section 4:} In total we find that the effective thermal conductivity significantly depends only on three of the four parameters, the relative thickness of the dust mantle and the thermal conductivities of core and mantle but the total size plays no major role for thermal conductivity. The thermal conductivity of the core is visible in the values of the effective thermal conductivity, i.e. by increasing the value of the mantle by a factor of two or more. However, the effective thermal conductivity is insensitive to variations of the core thermal conductivity in the range of reasonable values of $k_{dust}$. Assuming that the solid chondrules (core) have a fixed thermal conductivity on the order of $1\,\mathrm{W/m\,K}$, this leaves two variables in the context of photophoretic forces on chondrules, the thermal conductivity of the mantle and the dust mantle to core size ratio.

\section{Size sorting of chondrules}

The dependency of the effective thermal conductivity on the different parameters provides various options by which dust mantled chondrules can be sorted due to photophoresis. We only give a qualitative account here. Detailed treatments would require specific disk models. This would be a problem based on the results of this paper but otherwise being beyond its scope.

The sorting mechanism in \citet{Wurm2006} was based on a disk density profile which decreases with distance to the star. This way, the gas is supported against the stellar gravity and rotates slower than with Keplerian speed in order to rotate on a stable orbit. Small solid particles like chondrules embedded in the disk would couple to the gas rapidly, would rotate with the same slower orbital frequency but would not feel the support by the pressure gradient. Such particles drift inward \citep{Weidenschilling1977}. The particles are subject to a radial force 
\begin{equation}
F_D(a) = \frac{n\, R_{G}\,T_{gas} \,m_{par}}{\mu\, a} \; ,
\end{equation}
then which can be considered as residual gravity. Here $a$ is the radial distance to the star, $\mu$ is the molar mass of the gas, $n$ is the power of the radial distance dependency of the gas density ($n=11/4$ in the minimum mass solar nebula, see eq. \ref{eq::gasDensity} \citep{Hayashi1985}), $R_{G}$ is the gas constant and $m_{par}$ is the particle mass.
This includes gas drag in the way that the gas grain coupling time is much shorter than the orbital timescale and the problem reduces to a one dimensional radial problem.

\cite{Wurm2006} consider the simple photophoretic force estimate given by eq. \ref{forceA2}. and calculate the ratio 
\begin{equation}
\frac{F}{F_D}  \simeq \frac{\tilde{F}^{(2)}}{ F_D} \; .
\end{equation}
If this ratio is 1 inward and outward drift balance and chondrules are concentrated. This concentration distance depends on the strength of the photophoretic force. Therefore, chondrules with different photophoretic properties -- as outlined in this paper -- would concentrate at different locations.

\begin{figure} 
   \centering
   \def\svgwidth{\columnwidth} 
   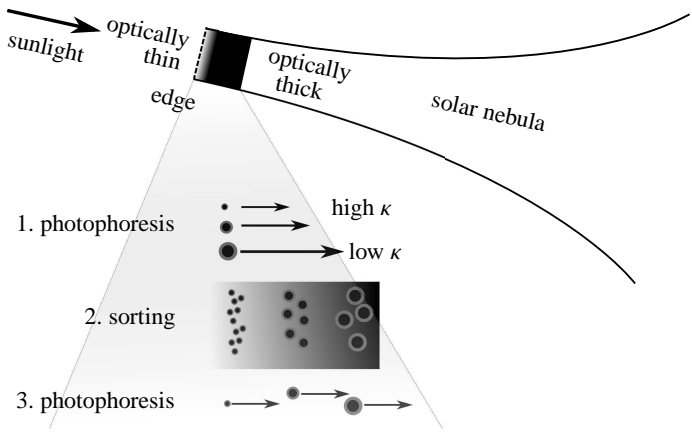
   \caption{Principles of sorting dust mantled chondrules by photophoresis}
   \label{fig10}
 \end{figure}

Particle sorting due to photophoresis might also take place at the inner edge of the disk, which might move over the asteroid belt \citep{Haack2007}. The idea is sketched in fig. \ref{fig10}. At the inner edge of a disk which is optically thick the sorting might work differently than described above. Per definition the optical depth decreases further away from the edge. Particles are therefore pushed outward as long as the decreasing photophoretic outward motion is more effective than any other inward directed effect like turbulent diffusion or any other remaining radial inward drift and the particles pushed more vigorously by photophoresis will move outward into the dark the furthest. Here, the sorting distances are strongly disk dependent but the principle is independent of the details.

\citet{Wurm2006} considered that particles with thermal conductivities varying by a factor 2 could easily be concentrated at different positions in a solar nebula almost $1\,\mathrm{AU}$ apart of each other. A rough linear scaling would imply that a difference in effective thermal conductivity on the order of 10\% would separate particles on the order of $0.1\,\mathrm{AU}$ or $15\times 10^6\,\mathrm{km}$. This is certainly sufficient to separate particles far enough to be incorporated into different parent bodies, eventually.


\textbf{Direct size sorting:} As calculated above the effective thermal conductivity does not change with total size. The photophoretic force has an intrinsic dependence on size though as $r^3$ (eq. \ref{forceNumerical}) with the correction factor $z^{(3)}$ included. For a homogeneous particle the gravitational pull by a star also depends on $r^3$ and the ratio of both forces is essentially independent of the total size. Without correction factor photophoresis would not sort by size directly. For the lowest thermal conductivities there is a size dependence in the correction factor which varies the photophoretic force by 10\% for the smallest and largest chondrule (see fig. \ref{fig5}). In that case direct size sorting of dust mantled chondrules with plausible mantle thickness by photophoresis is possible. It might also be considered here that the dust mantle has a different porosity as the core, which might vary by about a factor of 3. Nevertheless, for a constant size ratio between mantle and core a change in total size would not change the average density of the particle and this is not important. 

\textbf{Mantle to core size ratio sorting:} It is interesting to note that for a constant dust mantle to core size ratio no sorting due to the effective thermal conductivity would occur. However, considering the growth of dust mantles by accretion, \citet{Carballido2011} shows that the size ratio between mantle and core changes with overall size. This introduces an absolute size dependence of the effective thermal conductivities then: $\kappa\left(d_{dust}\left(\rho \right)  \right)$. Values estimated from fig. \ref{fig9} in \citet{Carballido2011} are e.g. a ~$0.3 \,\mathrm{mm}$ core with a ~$0.2\,\mathrm{mm}$ mantle ($d_{dust} = 0.6$) but a $1\,\mathrm{mm}$ core only has $0.4\,\mathrm{mm}$ mantle ($d_{dust} = 0.4$). Corresponding effective thermal conductivities from our simulations assuming the core to have, $k_{core} = 1 \,\mathrm{W/m\,K}$ and the rim to have $k_{dust} = 0.01\,\mathrm{W/m\,K}$ are $\kappa = 0.02 \,\mathrm{W/m\,K}$ and $\kappa = 0.03 \,\mathrm{W/m\,K}$, respectively. This is a difference of a factor $1.5$ which is certainly sufficient to separate these particles by photophoresis. A thinner mantle also corresponds to a slightly higher average mass density and gravitational attraction. This works in the same direction as decreasing the strength of the photophoretic force.

\textbf{Mantle thermal conductivity:} The exact thermal conductivity of the mantle is unknown as it depends on many parameters. Assuming that the material properties of the dust are the same on average, porosity of the dust rim is still a free parameter. The work by \citet{Bland2011} suggests that mantle porosities are on the order of $0.7$ or more. This is consistently on the same order of porosities to be expected from dust experiments \citep{Teiser2011}. For this porosity and micrometer size dust \citet{Krause2011} find thermal conductivities on the order of $0.01\,\mathrm{W/m\,K}$ and below. We assumed this value for dust mantles above. 
Typically smaller particles collide with dust at lower velocities than larger ones. The experiments by \citet{Teiser2011} and \citet{Kothe2011} show that such differences in collision velocities produce a difference in porosity. In that case the dust mantle porosity would systematically change (decrease) with total size. Typical differences are several percent. Assuming a porosity variation e.g. of $0.05$ would lead to a difference in thermal conductivity on the order of $1.5$ for the dust mantle \citep{Krause2011}. Varying the dust mantle thermal conductivity by this factor, e.g. between $0.01 \,\mathrm{W/m\,K}$  and $0.015 \,\mathrm{W/m\,K}$ for a core with $1 \,\mathrm{W/m\,K}$ changes the effective thermal conductivity by a similar factor of $1.54$ ($r_{core} = 1\,\mathrm{mm} , d_{dust} = 1$) or $1.46$ ($r_{core} = 1\,\mathrm{mm} , d_{dust} = 0.5$). This is on the order of the changes induced by the different mantle to core size ratios discussed in the section before. This provides the basis for a photophoretic size sorting. The trend would go in the same direction as the size ratio variations, i.e. larger particles have a lower effective thermal conductivity and would be found closer to the sun.

\textbf{Core thermal conductivity:} The third parameter influencing the effective thermal conductivity is the thermal conductivity of the core. As it is plausible that dust mantles have rather low thermal conductivity, the dependency on this parameter is not strong. As found in the simulations effective thermal conductivities for thin (20\% of core radius) dust mantles vary by 10\%. For more conductive rims of $k_{dust} = 0.1 \,\mathrm{W/m\,K}$ this might be somewhat larger but compared to the dependency on the size ratio between mantle and core the effect is small. We conclude that for core-mantle particles the thermal conductivity of the core will not be responsible for size sorting.

\citet{Wurm2006} estimate the drift velocity of chondrules due to photophoresis at $1 \,\mathrm{AU}$ which would be on the order of $0.1\,\mathrm{m/s}$. Separating the particles by $10 \times 10^6\,\mathrm{km}$ would require about $10^7\,\mathrm{s}$ or $1\,\mathrm{year}$ or $1$ orbit then which is fast compared to evolutionary time scales. If all mass is in chondrules and if chondrules do not evolve e.g. by growing to larger bodies then a maximum mass of separated chondrules can be estimated by the number of chondrules which would make the disk optically thick themselves. Assuming a disk height of $0.1 \,\mathrm{AU}$ at  $1 \,\mathrm{AU}$ distance to the star the area of the inner edge would be $2\pi\, \times 1 \,\mathrm{AU} \times 0.1 \,\mathrm{AU}$. Completely covered with chondrules of $1\,\mathrm{mm}$ diameter, the number of chondrules is $6.35 \times 10^{27}$ and assuming these chondrules have a mass density of $3\,\mathrm{g/cm^3}$ one chondrule has a mass of $1.6\,\mathrm{mg}$. The total number then is $10^{22}\,\mathrm{kg}$ which is about $10$ times the mass of Ceres. Being that large the estimate shows that the mechanism provides more than sufficient separation rates even if conditions are less perfect than assumed for this estimate.

\section{Caveats}\label{caveats}

\textbf{Radiation pressure}: Photophoresis is driven by radiation. Other movements of particles induced by radiation and more commonly considered are radiation pressure and the Poynting-Robertson effect \citep{Klacka2002}. The difference is that radiation pressure is based on the direct momentum transfer of photons while photophoresis is based on momentum transfer via gas molecules. For chondrule size particles and plausible gas densities photophoresis is much larger as seen in the following estimate.
Radiation pressure can be approximated by 
\begin{equation}
F_{rad}=\frac{I}{c}\pi r^2 \; .
\end{equation}
Compared to the classical photophoretic force estimate of eq. \ref{forceA2} which is sufficient for this estimate as photophoretic force it is
\begin{equation}
\tilde{F}^{(2)}/F_{rad} = \frac{c\,r\,p}{6\, k\,T_{gas}} \; .
\end{equation}
Taking a typical value of $k= 0.1 \,\mathrm{W/m\,K}$, a worst case of a small chondrule of $r=0.2\,\mathrm{mm}$ and a worst case temperature of $T_{gas}=280\, \mathrm{K}$ we get $F_{phot}/F_{rad} \simeq 357\, p$. The minimum mass solar nebula by \citet{Hayashi1985} has an ambient pressure of about $1.4 \,\mathrm{Pa}$ at $1 \,\mathrm{AU}$. Here, photophoresis is a factor of 500 stronger than radiation pressure. The pressure is $0.04 \,\mathrm{Pa}$ at $3\,\mathrm{AU}$ or distances related to the outer asteroid belt. Here, photophoresis is 14 times stronger than radiation pressure using the above conservative estimates. Therefore, photophoresis dominates as radiative force. 

\textbf{Rotation:}
Particle rotation was not considered here though it is important in the context of photophoresis. If particles would rotate more rapidly than the typical heat transfer time then the temperature gradient would not reach equilibrium and be reduced which leads to a decrease in the photophoretic force. We did not consider rotations for the following reasons. Any random rotation is damped on a time scale of the gas-grain coupling time. In the free molecular flow regime which is relevant for mm-particles in protoplanetary disks the coupling time is given as \citep{Blum1996} 
\begin{equation}
\tau_c = \epsilon\frac{m_{par}}{\sigma_{par}} \frac{1}{\rho_{gas} \bar{v}_{gas,th} }\;,
\end{equation}
with $\rho_{gas}$ being the gas density, $\sigma_{par}$ the particle's geometrical cross section, and the mean thermal velocity $\bar{v}_{gas,th}=\sqrt{\frac{8\, R_G T_{gas}}{\pi\, \mu}}$.
At $a=1 \,\mathrm{AU}$ in a minimum mass nebula it is \citep{Hayashi1985}
\begin{align}
      \rho_{gas}(a) &= 1.4 \cdot 10^{-6} \left(\frac{a}{1 \,\mathrm{AU}}\right)^{-11/4}  \,\mathrm{kg/m^3} \;, \label{eq::gasDensity}\\
      T_{gas}(a) &= 280\left( \frac{a}{1 \,\mathrm{AU}}\right) ^{-1/2}\,\mathrm{K}\;,\\
      p(a) &= 3551\,\rho_{gas}(a) \, T_{gas}(a)\,\mathrm{\frac{m}{K\,s^2}}\;.
\end{align}
Then, $\tau_c$ is on the order of $10^3\,\mathrm{s}$ for a $1\,\mathrm{mm}$ particle. Random rotation is therefore decaying rapidly. Random rotations would only be important in decreasing the photophoretic strength effectively if they would continuously be excited on similar time scales. A way to generate particle rotation is inter-particle collisions. \citet{Krauss2007} estimated that the time between collisions of mm-size particles of mass-density $\rho_{par}$ in a `typical' disk is given by 
\begin{equation}
\tau_{coll}=33\frac{\rho_{par}}{\rho_{gas}} \frac{r}{v_{par,rel}}\;,
\end{equation}
which then is about a factor of $10^4$ longer than the gas grain coupling time ($v_{par,rel}=1\,\mathrm{m/s})$. Therefore, even assuming strongly varying disk conditions, rotation is damped much more rapidly and the total time of potentially decreased photophoretic strength is not significant.

Another source of random rotation to consider is Brownian motion. The thermal rotation energy of a spherical particle is given by $E_{rot}=\frac{3}{2}k_B T$, which leads to
\begin{equation}
\omega = \sqrt{\frac{3 k_B T_{gas}}{J_{par}}}=\sqrt{\frac{45 k_B T_{gas}}{8 \pi r^5 \, \rho_{par}}}
\end{equation}
with $J_{par}$ being the sphere's moment of inertia. To rotate by $180^{\circ}$ and change front and backside it is
\begin{equation}
\tau_{rot(180^{\circ})} = \sqrt{\frac{8 \pi^3 r^5 \, \rho_{par}}{45 k_B T_{gas}}} \;.
\end{equation}
For chondrules this is on the order of $10^4\,\mathrm{s}$. The time scale for thermal conduction through the particle is given by
\begin{equation}
\tau_{cond} = \frac{r^2}{k}\rho_{par}\,c_{par} \;.
\end{equation}
Here, $c_{par}$ is the particle's heat capacity at constant pressure. For chondrules $\tau_{cond}$ is on the order of $1\,\mathrm{s}$. Therefore, a temperature gradient reestablishes itself much faster than thermal rotation can disalign it and hence Brownian Rotation can be neglected, too.

Besides random rotation exciting a continuous rotation is possible by photophoresis itself. As particles are never ideal spheres torques are induced. One can argue though that only torques around the direction of light will induce a systematic rotation around the axis aligned to the direction of light. Any other torque will only lead to the stable alignment of the particle. To prove this \citet{vanEymeren2012} carried out a detailed study of particle rotation. In their setup particles were levitated by means of thermophoresis and photophoresis \citep{Kelling2011}. \citet{vanEymeren2012} found that at least more than 95\% of all particles rotate around the vertical direction which coincides with the direction of illumination. Such a rotation does not change front and back with respect to the illumination and is not important for the photophoretic strength along the direction of light. For this reason rotation is no issue in particle transport by stellar radiation. It might be noted that the situation is different for radiation pressure acting in the current solar system. However, here no gas and therefore no damping is present. 

\textbf{Heat exchange with the ambient gas:}
In our heat transfer calculations we did not include heat exchange with the ambient gas. Heat transfer to the ambient gas for large Knudsen numbers can be quantified as \citep{Stoffels1996}

\begin{equation}
I_{cond}=\frac{\frac{c_p}{c_v}+1}{8\left( \frac{c_p}{c_v}-1\right) }\alpha \Delta T p \sqrt{\frac{2\,k_B}{\pi\,T_{gas}m_{gas}}}\;,
\end{equation}
where $I$ is the heat flux (power per surface area).

In the case of molecular hydrogen, which consists of two atoms, $c_p/c_v$  can by approximated by $9/7$. $\Delta T$ denotes the temperature difference between the chondrule and the surrounding gas. For a chondrule at $2 \,\mathrm{AU}$ in a minimum mass nebula by \citet{Hayashi1985} we have a radiative flux $I=341\,\mathrm{W/m^2},\, p= 0.15\, \mathrm{Pa}, T_{gas}=262\,  \mathrm{K}$. Maximum temperature differences are $\Delta T = 10\, \mathrm K$. This leads to $I_{cond}=35\,\mathrm{W/m^2}$.
This is an order of magnitude less than the incoming radiative flux and can be neglected in the context
of photophoretic sorting of chondrules in protoplanetary disks.

Typical laboratory conditions are $I=20\,\mathrm{kW/m^2}$ (maximum $\Delta T=400\, \mathrm{K}$, see fig. \ref{fig2}), $p= 1\,\mathrm{Pa}$, $m_{gas}=29\,\mathrm{u}$ (air), and $T_{gas}=293\,\mathrm{K}$ which yields $I_{cond}=315\,\mathrm{W/m^2}$. This is also much lower than the incoming light flux and cooling by gas interaction still can be neglected. In case of higher pressures ($p= 10\,\mathrm{Pa}\dots 100\,\mathrm{Pa}$) heat exchange with the gas becomes important, but at the same time the assumption of photophoresis in the free flow regime is also no longer valid. For transitional disks assumed to be of low pressure our treatment holds. On the other hand higher disk (gas) densities lead to a reduced light flux as they are usually assumed to go with higher absorption. This also changes the photophoretic strength and transport. In the extreme case of an optically thick disk, photophoresis does not work. Therefore, high density regions of a disk require an extended treatment of photophoresis, but this is beyond the scope of this paper.

\section{Conclusion}

In summary there are two options to size sort chondrules with dust mantles by photophoresis. The mantle thickness variation with size \citep{Carballido2011} naturally induces different effective thermal conductivities. Plausible porosity variations with size act in the same direction and on the same magnitude. Therefore, if dust covered chondrules are subject to a directed radiation they likely get size sorted by photophoresis. 
In the original work by \citet{Wurm2006} the solar nebula was assumed to be optically thin throughout. This is not likely to be the case for the very young nebula. However, an increasing number of transition disks is found, which have an optical thin inner clearing up to tens of $\mathrm{AU}$ \citep{Najita2007, Sicilia-Aguilar2008}. Some of these show active signs of accretion and therefore the optical thin part is still gaseous. Within such a clearing the idea by \citet{Wurm2006} can naturally be applied. The effective thermal conductivities of dust mantled chondrules are 1 to 2 orders of magnitude lower than the thermal conductivity of bare chondrules assumed by \citet{Wurm2006}. This increases the photophoretic strength by a similar factor. Therefore, dust mantled chondrules can concentrate in the asteroid belt region even if the inner clearing has a significantly reduced gas density. Creating the inner clearing takes some time. This would also fit well with the fact that chondrules are forming their parent bodies only a few million years after the beginning of the solar nebula.
\citet{Haack2007} also considered the idea that the inner edge of the disk might be important. This would shift the focus from the spatially extended inner clearing to the actual edge that is moving outwards. Here the disk is supposed to be of its original density. Particles exposed to the starlight at the edge would feel a photophoretic force and be pushed outwards until the optical depth is too large. The final position with respect to the edge will depend on the susceptibility to photophoresis and particles with lower effective thermal conductivity will move further out. This will introduce a size sorting at the edge of the disk as it moves inside out. If parent bodies form in this increased density environment chondrules will be size sorted. A detailed treatment of this is beyond the scope of this paper as it would have to focus on disk models, while this work is dedicated to the photophoretic properties needed as input.
Here, we evaluated the effective thermal conductivity of a dust mantled chondrule in the context of photophoretic forces. It strongly depends on the size ratio between mantle and core and the thermal conductivity of the mantle. Based on these results a size sorting of dust mantle chondrules by photophoresis is possible.

\begin{acknowledgements}
This work is funded by the Deutsche Forschungsgemeinschaft within the priority program SPP 1385. We also thank the referee for a very constructive review of the manuscript.
\end{acknowledgements}

\bibliographystyle{aa} 
\bibliography{Referenzen} 

\end{document}